\documentclass[12pt,journal,epsfig,onecolumn,peerreviewca,draftclsnofoot]{IEEEtran}

\usepackage[dvips]{graphicx}
\usepackage{graphicx}
\usepackage{amssymb}
\usepackage{cite}
\usepackage{subfigure}
\usepackage{amsmath}

\usepackage{algorithm}
\usepackage{algorithmic}

\usepackage{tikz}
\usepackage{lipsum}
\IEEEoverridecommandlockouts

\newcommand\copyrighttext{%
  \footnotesize \textcopyright 2020 IEEE. Personal use of this material is permitted.
  Permission from IEEE must be obtained for all other uses, in any current or future
  media, including reprinting/republishing this material for advertising or promotional
  purposes, creating new collective works, for resale or redistribution to servers or
  lists, or reuse of any copyrighted component of this work in other works.
  }
\newcommand\copyrightnotice{%
\begin{tikzpicture}[remember picture,overlay]
\node[anchor=south,yshift=10pt] at (current page.south) {\fbox{\parbox{\dimexpr\textwidth-\fboxsep-\fboxrule\relax}{\copyrighttext}}};
\end{tikzpicture}%
}

\begin{document}

\title{Joint Transceiver and Large
Intelligent Surface Design for Massive MIMO MmWave Systems}

\author{Peilan Wang, Jun Fang, Linglong Dai, and Hongbin Li, ~\IEEEmembership{Fellow,~IEEE}
\thanks{Peilan Wang and Jun Fang are with the National Key Laboratory
of Science and Technology on Communications, University of
Electronic Science and Technology of China, Chengdu 611731, China,
Email: JunFang@uestc.edu.cn}
\thanks{Linglong Dai is with the Department of Electronic Engineering, Tsinghua
University, Beijing 100084, China, E-mail: daill@tsinghua.edu.cn}
\thanks{Hongbin Li is with the Department of Electrical and Computer Engineering,
Stevens Institute of Technology, Hoboken, NJ 07030, USA, E-mail:
Hongbin.Li@stevens.edu}
\thanks{This work was supported in part by the National Science
Foundation of China under Grant 61522104.}}

\maketitle

\copyrightnotice

\begin{abstract}
Large intelligent surface (LIS) has recently emerged as a
potential low-cost solution to reshape the wireless propagation
environment for improving the spectral efficiency. In this paper,
we consider a downlink millimeter-wave (mmWave)
multiple-input-multiple-output (MIMO) system, where an LIS is
deployed to assist the downlink data transmission from a base
station (BS) to a user equipment (UE). Both the BS and the UE are
equipped with a large number of antennas, and a hybrid
analog/digital precoding/combining structure is used to reduce the
hardware cost and energy consumption. We aim to maximize the
spectral efficiency by jointly optimizing the LIS's reflection
coefficients and the hybrid precoder (combiner) at the BS (UE). To
tackle this non-convex problem, we reformulate the complex
optimization problem into a much more friendly optimization
problem by exploiting the inherent structure of the effective
(cascade) mmWave channel. A manifold optimization (MO)-based
algorithm is then developed. Simulation results show that by
carefully devising LIS's reflection coefficients, our proposed
method can help realize a favorable propagation environment with a
small channel matrix condition number. Besides, it can achieve a
performance comparable to those of state-of-the-art algorithms,
while at a much lower computational complexity.
\end{abstract}


\section{Introduction}
Millimeter-wave (mmWave) communication is regarded as a promising
technology for future cellular networks due to its large available
bandwidth and the potential to offer gigabits-per-second
communication data rates
\cite{RappaportMurdock11,RanganRappaport14,GhoshThomas14,SunRappaort18}.
Utilizing large antenna arrays is essential for mmWave systems as
mmWave signals incur a much higher free-space path loss compared
to microwave signals below 6 GHz
\cite{AlkhateebMo14,SwindlehurstAyanoglu14}. Large antenna arrays
can help form directional beams to compensate for severe path
loss incurred by mmWave signals. On the other hand, high
directivity makes mmWave communications vulnerable to blockage
events, which can be frequent in indoor and dense urban
environments. For instance, due to the narrow beamwidth of mmWave
signals, a very small obstacle, such as a person's arm, can
effectively block the link \cite{AbariBharadia17,Maccartney17}.

To address this blockage issue, large intelligent surface has
been recently introduced to improve the spectral efficiency and
coverage of mmWave systems
\cite{WangFang19,TanSun18,StefanRenzo19,YangWen20}. Large intelligent
surface (LIS), also referred to intelligent reflecting surface
(IRS), has emerged as a promising technology to realize a smart
and programmable wireless propagation environment via
software-controlled reflection
\cite{LiaskosNie18,LiangLong19,HuRusek18,BasarDiRenzo19,WuZhang20,HanTang19}.
Specifically, LIS is a two-dimensional artificial structure,
consisting of a large number of low-cost, passive, reconfigurable
reflecting elements. With the help of a smart controller, the
incident signal can be reflected with reconfigurable phase shifts
and amplitudes. By properly adjusting the phase shifts, the
LIS-assisted communications system can achieve desired
properties, such as extending signal coverage
\cite{WuZhang18}, improving energy
efficiency\cite{HuangZappone19}, mitigating interference
\cite{WuZhang19a,LiFang19}, enhancing system security
\cite{CuiZhang19}, and so on.

A key problem of interest in LIS-assisted communication systems is
to jointly devise the reflection coefficients at the LIS and the
active precoding matrix at the BS to optimize the system
performance. A plethora of studies have been conducted to
investigate the problem under different system setups and
assumptions, e.g.
\cite{WuZhang18,HuangZappone19,WuZhang19a,WangFang19,LiFang19,
CuiZhang19,NingChen20,ZhangZhang19,StefanRenzo19,YingGao20,PanRen19,PanRen19b}.
Among them, most focused on \emph{single-input-single-output}
(SISO) or \emph{multiple-input-single-output} (MISO) systems
\cite{WuZhang18,HuangZappone19,WuZhang19a,WangFang19,LiFang19,CuiZhang19}.
For the MIMO scenario where both the BS and the UE are equipped
with multiple antennas, \cite{NingChen20} proposed to optimize the
spectral efficiency via maximizing the Frobenius-norm of the
effective channel from the BS to the UE. Nevertheless, maximizing
the Frobenius-norm of the effective channel usually results in a
large condition number, and as a result, its performance
improvement is limited. In \cite{ZhangZhang19}, an alternating
optimization (AO)-based method was developed to maximize the
capacity of LIS-aided point-to-point MIMO systems via jointly
optimizing LIS's reflection coefficients and the MIMO transmit
covariance matrix. Specifically, the AO-based method sequentially
and iteratively optimizes one reflection coefficient at a time by
fixing the other reflection coefficients. Although achieving
superior performance, this sequential optimization incurs an
excessively high computational complexity for practical systems.
In \cite{StefanRenzo19}, two optimization schemes were proposed to
enhance the channel capacity for LIS-assisted mmWave MIMO systems.
Nevertheless, this work is confined to the scenario where only a
single data stream is transmitted from the BS to the UE. To
improve the bit error rate (BER) performance, \cite{YingGao20}
proposed a geometric mean decomposition (GMD)-based beamforming
scheme for LIS-assisted mmWave MIMO systems. To simplify the
problem, only the angle of arrival (AoA) of the line-of-sight
(LOS) BS-LIS link and the angle of departure (AoD) of the LOS
LIS-UE path were utilized to design the reflection coefficients.
This simplification, however, comes at the cost of sacrificing the
spectral efficiency. Besides the above works, other studies
\cite{PanRen19,PanRen19b} considered multiple-user (MU)-MIMO
scenarios. For example, \cite{PanRen19} considered using the LIS
at the cell boundary to assist downlink transmission to cell-edge
users.

In this paper, we consider a point-to-point LIS-assisted mmWave
systems with large-scale antenna arrays at the transmitter and
receiver, where an LIS is deployed to assist the downlink
transmission from the transmitter (i.e. BS) to the receiver (i.e.
UE). Our objective is to maximize the spectral efficiency of the
LIS-assisted mmWave system by jointly optimizing the LIS
reflection coefficients and the hybrid precoding/combining
matrices associated with the BS and the UE. To address this
non-convex optimization problem, we first decouple the LIS design
from hybrid precoder/combiner design. We then focus on the
reflection coefficient optimization problem. By exploiting the
inherent structure of the effective mmWave channel, it is shown
that this complex LIS optimization problem can be reformulated
into an amiable problem that maximizes the spectral efficiency
with respect to the passive beamforming gains (which have an
explicit expression of the reflection coefficients) associated
with the BS-LIS-UE composite paths. A manifold-based optimization
method is then developed to solve the LIS (also referred to as
passive beamforming) design problem. Simulation results show that
our proposed method can help create a favorable propagation
environment with a small channel matrix condition number. Also,
the proposed method exhibits a significant performance improvement
over the sum-path-gain maximization method \cite{NingChen20} and
achieves a performance similar to that of the AO-based method
\cite{ZhangZhang19}, while at a much lower computational
complexity.

%where the objective is to devise the LIS's reflection coefficients
%such that the largest few singular values of the effective channel
%can be well-shaped to maximize the spectral efficiency

%A key insight here is to realize that the singular values of the
%effective channel can be conveniently approximated by the complex
%gains associated with the BS-LIS-UE composite paths.

The rest of the paper is organized as follows. In Section
\ref{sec-system}, the system model and the joint active and
passive beamforming problem are discussed. The passive beamforming
problem is studied in Section \ref{sec-LIS-BF}, where a
manifold-based optimization method is developed. The hybrid
precoding/combining design problem is considered in Section
\ref{sec-hyb-man}. The comparison with the state-of-art algorithms
is discussed in Section \ref{sec-comp}. Simulation results are
provided in Section \ref{sec-simu}, followed by concluding remarks
in Section \ref{sec-conclu}.

\begin{figure*}
    \centering
    {\includegraphics[width=6in]{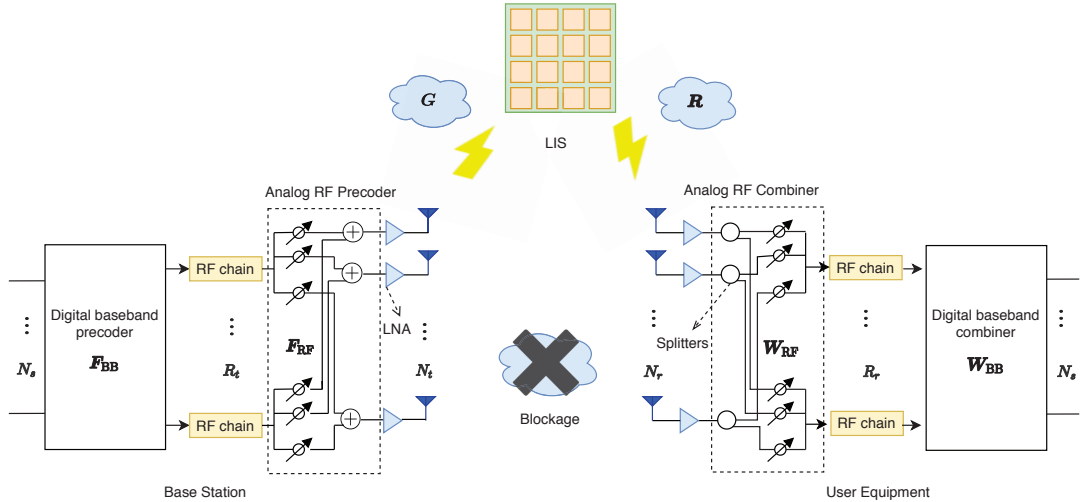}}
    \caption{LIS-assisted mmWave hybrid downlink system.} \label{fig1}
\end{figure*}

\section{System Model and Problem Formulation} \label{sec-system}
\subsection{System Model}
We consider a point-to-point mmWave MIMO system as illustrated in
Fig. \ref{fig1}, where the BS and the UE are equipped with a large
number of antennas, and the LIS is deployed to assist the downlink
data transmission from the BS to the UE. For simplicity, we assume
that the direct link between the BS and the UE is blocked due to
unfavorable propagation conditions. Nevertheless, our proposed
scheme can be extended to the scenario in which a direct link between
the BS and the UE is available. The BS is equipped with $N_t$ antennas and
$R_t$ radio frequency (RF) chains, and the UE is equipped with
$N_r$ antennas and $R_r$ RF chains, where $R_t\ll N_t$ and $R_r\ll
N_r$. To exploit the channel diversity, multiple, say $N_s$, data
streams are simultaneously sent from the BS to the UE, where
$N_s\leq \min \{R_{t}, R_{r}\}$. At the BS, a digital baseband
precoder $\boldsymbol{F}_{\text{BB}} \in \mathbb{C}^{R_{t}\times
N_s}$ is first applied to the transmitted signal
$\boldsymbol{s}\in \mathbb C^{N_s}$, then followed by an analog RF
beamformer $\boldsymbol{F}_{\text{RF}} \in \mathbb{C}^{N_{t}
\times R_t}$. The transmitted signal can be written as
\begin{align}
\boldsymbol{x} = \boldsymbol{F}_{\text{RF}}
\boldsymbol{F}_{\text{BB}} \boldsymbol{s},
\end{align}
where $\boldsymbol{s}$ is assumed to satisfy $\mathbb
E[\boldsymbol{s} \boldsymbol{s}^H] =\boldsymbol{I}$. Also, a
transmit power constraint is placed on the hybrid precoding matrix
$\boldsymbol{F}_{\text{RF}} \boldsymbol{F}_{\text{BB}}$, i.e. $\|
\boldsymbol{F}_{\text{RF}} \boldsymbol{F}_{\text{BB}}\|_F^2 \leq
\rho$, $\rho$ is the transmit power.

The transmitted signal arrives at the UE via propagating through
the BS-LIS-UE channel, where the LIS comprises $M$ passive
reflecting elements and each element behaves like a single
physical point which combines all the received signals and then
re-scatters the combined signal with an adjustable phase shift
\cite{WuZhang19a}. Let $\boldsymbol{G} \in \mathbb C^{M \times
N_t}$ denote the channel from the BS to the LIS, and
$\boldsymbol{R} \in \mathbb C^{N_r \times M}$ denote the LIS-UE
channel. Also, let $\phi_m \in [0,2\pi]$ denote the phase shift
introduced by the $m$th element of the LIS, and
\begin{align}
    \boldsymbol{\Phi} \triangleq {\rm diag}(e^{j\phi_1},
    \cdots, e^{j \phi_M}) \in \mathbb C^{M \times M}.
\end{align}
The signal received by the UE can thus be given as
\begin{align}
    \boldsymbol{y}  = \boldsymbol{W}_{\text{BB}}^H
    \boldsymbol{W}_{\text{RF}}^H  \boldsymbol{H}_{\rm eff}
    \boldsymbol{F}_{\text{RF}} \boldsymbol{F}_{\text{BB}}\boldsymbol{s}
    + \boldsymbol{W}_{\text{BB}}^H \boldsymbol{W}_{\text{RF}}^H  \boldsymbol{n},
\end{align}
where $\boldsymbol{W}_{\text{RF}} \in \mathbb{C}^{N_{r} \times R_r}$ and
$\boldsymbol{W}_{\text{BB}} \in \mathbb{C}^{R_{r} \times N_s}$
represent the analog combiner and the digital baseband combiner,
respectively, $\boldsymbol{H}_{\rm eff} \triangleq \boldsymbol{R}
\boldsymbol{\Phi} \boldsymbol{G}$ stands for the effective (i.e.
cascade) channel from the BS to the UE, and $\boldsymbol{n} \in \mathbb C^{N_r}
\sim \mathcal{CN}(0,\sigma^2 \boldsymbol{I})$
denotes the additive white Gaussian noise.

In this paper, we assume that perfect channel state information
(CSI) is known for joint transceiver and LIS design. Channel
estimation for LIS-assisted systems is an important and
challenging issue that has been studied in a variety of works,
e.g. \cite{DeepakHaakan19,WangFang19b,ChenLiang19,LinYu20}. In particular,
\cite{WangFang19b,ChenLiang19,LinYu20} discussed how to estimate the
channel for LIS-assisted mmWave systems. Suppose that the
transmitted signal follows a Gaussian distribution. The achievable
spectral efficiency can be expressed as \cite{YuShen16}
\begin{align}
R = \log_2 {\rm det} \bigg(  \boldsymbol{I} + & \frac{
1}{\sigma^2} ( \boldsymbol{W}_{\text{RF}}
\boldsymbol{W}_{\text{BB}}) ^{\dagger} \boldsymbol{H}_{ \rm {eff}
}  \boldsymbol{F}_{\text{RF}} \boldsymbol{F}_{\text{BB}}
\nonumber \\
& \times \boldsymbol{F}_{\text{BB}}^H \boldsymbol{F}_{\text{RF}}^H
\boldsymbol{H}_{  \rm eff}^H      ( \boldsymbol{W}_{\text{RF}}
\boldsymbol{W}_{\text{BB}}) \bigg). \label{spectral-efficiency}
\end{align}
where $\boldsymbol{A}^{\dagger}$ denotes the Moore-Penrose pseudo inverse of the matrix $\boldsymbol{A}$.
It should be noted that since the analog precoder and combiner are
implemented by analog phase shifters, entries of
$\boldsymbol{F}_{\text{RF}}$ and $\boldsymbol{W}_{\text{RF}}$ have
constant modulus.

\subsection{Channel Model}
Due to the small wavelength, mmWave has limited ability to
diffract around obstacles. As a result, mmWave channels exhibit a
sparse multipath structure and are usually characterized by the
Saleh-Valenzuela (SV) channel model
\cite{SawadaShoji06,AyachRajagopal14,GaoDai16,AkdenizLiu14,YingGao20}.
Suppose uniform linear arrays (ULAs) are employed at the BS and
the UE, and the LIS is a uniform planar array (UPA) consisting of
a large number of reconfigurable passive elements. The BS-LIS
channel $\boldsymbol{G}$ and the LIS-UE channel $\boldsymbol{R}$
can be given as
\begin{align}
\boldsymbol{G} = &\sqrt{ \frac{N_tM}{P}}  \sum_{i=1}^{P}
\tilde\alpha_i \boldsymbol{a}_{R_{   \rm  LIS}}(\theta_i^r,
\eta_i^r) \boldsymbol{a}_{T_{  \rm   BS}}^H(\gamma_i^t),
\label{BS-IRS}
\\
\boldsymbol{R} = & \sqrt{ \frac{MN_r}{L}}  \sum_{i=1}^{L} \tilde
\beta_i \boldsymbol{a}_{R_{  \rm   UE}} (\gamma_i^r)
\boldsymbol{a}_{T_{ \rm   LIS}}^H(\theta_i^t, \eta_i^t),
\label{IRS-UE}
\end{align}
where $P$ ($L$) is the total number of signal paths between the BS
and the LIS (the LIS and the UE), $\theta_{i}^{r}$
($\theta_i^{t}$) and $\eta_{i}^{r}$ ($\eta_i^{t}$) denote the
azimuth and elevation angles of arrival (departure) associated
with the LIS, $\gamma_i^{r}$ ($\gamma_i^{t}$) represents the angle
of arrival (departure) associated with the UE (BS),
$\tilde\alpha_i$ $(\tilde \beta_i)$ is the complex channel gain,
$\boldsymbol{a}_{R_{j}}, j \in \{ {  \rm  LIS, UE}\}$ and
$\boldsymbol{a}_{T_{i}}, i \in \{ {\rm BS, LIS}\}$ denote the
normalized array response vectors associated with the receiver and
the transmitter, respectively. Specifically, for ULA with $N$
antennas, the normalized array response vector is given by
\begin{align}
\boldsymbol{a}(\gamma) = \frac{1}{\sqrt{N}} [1\phantom{0}
\cdots\phantom{0}e^{j \frac{2\pi d}{\lambda}
(n-1)\sin(\gamma)}\phantom{0} \cdots\phantom{0}  e^{j \frac{2\pi
d}{\lambda} (N-1)\sin(\gamma)}]^T,
\end{align}
where $d$ and $\lambda$ are the antenna spacing and the signal
wavelength. For UPA with $M = M_y \times M_z$ elements, the
normalized array response vector can be written as
\begin{align}
\boldsymbol{a}(\theta, \eta) = \frac{1}{\sqrt{M}}[1\phantom{0}
\cdots\phantom{0} e^{j \frac{2 \pi d}{\lambda}
( (m_1 -1)\cos(\eta) \sin(\theta) + (m_2 -1) \sin(\eta))} \nonumber \\
\cdots\phantom{0} e^{j \frac{2 \pi d}{\lambda} ((M_y -1)\cos(\eta)
\sin(\theta) + (M_z -1) \sin(\eta))}]^T.
\end{align}

\subsection{Problem Formulation}
Our objective is to jointly devise the hybrid precoding/combining
matrices and the passive beamforming matrix $\boldsymbol{\Phi}$ to
maximize the spectral efficiency (\ref{spectral-efficiency}):
\begin{align}
\max_{\{\boldsymbol{F}_{\text{RF}},\boldsymbol{F}_{\text{BB}},\boldsymbol{W}_{\text{RF}},
\boldsymbol{W}_{\text{BB}},\boldsymbol{\Phi}\}} \quad & \log_2
{\rm det} \big( \boldsymbol{I} + \frac{ 1}{ \sigma^2} (
\boldsymbol{W}_{\text{RF}} \boldsymbol{W}_{\text{BB}} ) ^{\dagger}
\boldsymbol{H}_{ \rm {eff} }
\nonumber \\
& \times \boldsymbol{F}_{\text{RF}}
\boldsymbol{F}_{\text{BB}}\boldsymbol{F}_{\text{BB}}^H
\boldsymbol{F}_{\text{RF}}^H \boldsymbol{H}_{  \rm eff}^H      (
\boldsymbol{W}_{\text{RF}}\boldsymbol{W}_{\text{BB}}
) \big) \nonumber\\
\text{s.t.}\quad  &\| \boldsymbol{F}_{\text{RF}}
\boldsymbol{F}_{\text{BB}}\|_F^2 \leq \rho, \nonumber\\
& |\boldsymbol{F}_{\text{RF}}(i,j)| =
|\boldsymbol{W}_{\text{RF}}(i,j)|=1, \quad \forall i, j, \nonumber \\
&
\boldsymbol{H}_{\rm{eff}}=\boldsymbol{R}\boldsymbol{\Phi}\boldsymbol{G}
\nonumber\\
&\boldsymbol{\Phi} = {\rm diag}(e^{j\phi_1}, \cdots, e^{j
\phi_M}). \label{opt1}
\end{align}
Such an optimization problem is challenging due to the
non-convexity of the objective function and the per-element unit-modulus constraint
placed on analog precoding and combining matrices.

%we consider a alternating optimization scheme to decouple the
%phase shift matrix $\boldsymbol{\Phi}$ and the hybrid
%precoding/combining matrix. Specifically, we investigate the
%inherent relationship between the spectral efficiency and the
%phase shift matrix $\boldsymbol{\Phi}$ by adopting the
%near-optimal hybrid precoding/combining design. Hence, we optimize
%$\boldsymbol{\Phi}$ firstly and then design the hybrid
%precoding/combining matrix with the optimized $\boldsymbol{\Phi}$.

To simplify the problem, we first ignore the constraint introduced
by the hybrid analog/digital structure and consider a fully
digital precoder/combiner. Let
$\boldsymbol{F}\in\mathbb{C}^{N_t\times N_s}$ be a fully digital
precoder which has the same size as the hybrid precoding matrix
$\boldsymbol{F}_{\text{RF}}\boldsymbol{F}_{\text{BB}}$, and
$\boldsymbol{W}\in\mathbb{C}^{N_r\times N_s}$ be a fully digital
combiner which has the same size as the hybrid combining matrix
$\boldsymbol{W}_{\text{RF}}\boldsymbol{W}_{\text{BB}}$. The
problem (\ref{opt1}) can be simplified as
\begin{align}
\max_{\{\boldsymbol{F},\boldsymbol{W},\boldsymbol{\Phi}\}} \quad &
\log_2 {\rm det} \big( \boldsymbol{I} + \frac{ 1}{
\sigma^2} (\boldsymbol{W}) ^{\dagger} \boldsymbol{H}_{ \rm {eff} }
 \boldsymbol{F}\boldsymbol{F}^H \boldsymbol{H}_{  \rm
eff}^H  \boldsymbol{W}
\big) \nonumber\\
\text{s.t.}\quad  &\| \boldsymbol{F}\|_F^2 \leq \rho \nonumber \\
&
\boldsymbol{H}_{\rm{eff}}=\boldsymbol{R}\boldsymbol{\Phi}\boldsymbol{G}
\nonumber\\
& \boldsymbol{\Phi} = {\rm diag}(e^{j\phi_1},
    \cdots, e^{j \phi_M}).
 \label{opt2}
\end{align}
Once an optimal fully digital precoder/combiner is obtained, we
can employ the manifold optimization-based method \cite{Kasai18}
to search for a hybrid precoding/combining matrix to approximate
the optimal fully digital precoder/combiner. Such a strategy can
be well justified because it has been shown in many previous
studies \cite{AyachRajagopal14,Kasai18,YuShen16,GaoDai16} that, due to
the sparse scattering nature of mmWave
channels, hybrid beamforming/combining with a small number of RF
chains can asymptotically approach the performance of fully digital
beamforming/combining for a sufficiently large number of antennas at the transceiver.

Given a fixed $\boldsymbol{\Phi}$ (i.e.
$\boldsymbol{H}_{\rm{eff}}$), the optimal $\boldsymbol{F}$ and
$\boldsymbol{W}$ can be obtained via the singular value
decomposition (SVD) of $\boldsymbol{H}_{\rm eff}$. Define the
effective channel's ordered SVD as
\begin{align}
\boldsymbol{H}_{\rm   eff} = \boldsymbol{U} \boldsymbol{\Sigma}
\boldsymbol{V}^H =
\begin{matrix}
[ \boldsymbol{U}_1 & \boldsymbol{U}_2 ] \end{matrix} \left[
\begin{matrix}
\boldsymbol{\Sigma}_1 & \boldsymbol{0} \\ \boldsymbol{0}
&\boldsymbol{\Sigma}_2\end{matrix} \right] \left[
\begin{matrix}
\boldsymbol{V}_1^H  \\  \boldsymbol{V}_2^H    \end{matrix} \right],
\end{align}
where $\boldsymbol{U}$ is an $N_r \times Q$ unitary matrix,
$\boldsymbol{\Sigma}$ is a $Q \times Q$ diagonal matrix of
singular values, $\boldsymbol{V}$ is an $N_t \times Q$ unitary
matrix, in which $Q\triangleq{ \rm rank} (\boldsymbol{H}_{ \rm
eff})$. The matrix $\boldsymbol{\Sigma}_1$ is of dimension $N_s
\times N_s$ and $\boldsymbol{V}_1$ is of dimension $N_t \times
N_s$. Then with a fixed $\boldsymbol{\Phi}$, the optimal solution
to (\ref{opt2}) is given as
\begin{align}
\boldsymbol{F}_{\text{opt}} = \boldsymbol{V}_1
\boldsymbol{\Lambda}^{1/2}  ,\quad \boldsymbol{W}_{\text{opt}} =
\boldsymbol{U}_1, \label{digi-opt}
\end{align}
where $\boldsymbol{\Lambda} = {\rm diag}(p_1,\ldots, p_{N_s})$,
$p_i= {\rm max} (1/\lambda- \sigma^2/
\boldsymbol{\Sigma}_1^2(i,i),0), i = 1, \ldots, N_s$ denotes the
optimal amount of power allocated to the $i$th data stream,
$1/\lambda$ is the water level satisfying $\sum_{i=1}^{N_s} p_i=
\rho$. Thanks to the massive array gain provided by the LIS and
large number of antennas at the BS, the effective signal-to-noise
ratio (SNR) is large, in which case an equal power allocation
scheme is near-optimal \cite{AyachRajagopal14}. Therefore we can
approximate $\boldsymbol{F}_{\text{opt}}$ as:
\begin{align}
\boldsymbol{F}_{\text{opt}}\approx\sqrt{\frac{{\rho}}{N_s}
}\boldsymbol{V}_1.
\end{align}

Substituting the optimal fully digital precoder and combiner into
(\ref{opt2}), we arrive at a problem which concerns only the
optimization of the passive beamforming matrix
$\boldsymbol{\Phi}$:
\begin{align}
\max_{ \boldsymbol{\Phi}} \quad &   \log_2 {\rm det} \left(
\boldsymbol{I} + \frac{ \rho}{N_s \sigma^2} \boldsymbol{\Sigma}_1^2 \right) \nonumber \\
 {\rm s.t.} \quad  & \boldsymbol{\Phi} = {\rm diag}(e^{j\phi_1},
    \cdots, e^{j \phi_M}).
\label{opt3}
\end{align}
The above objective function can be bounded by
\begin{align}
\log_2 {\rm det} \left( \boldsymbol{I} + \frac{\rho}{N_s \sigma^2}
\boldsymbol{\Sigma}_1^2 \right) \stackrel{(a)} \leq & N_s
\log_2(1+
\frac{\rho}{N_s^2 \sigma^2}  {\rm tr}( \boldsymbol{\Sigma}_1^2) ) \nonumber \\
 \stackrel{(b)}\leq & N_s\log_2(1+  \frac{\rho}{N_s^2 \sigma^2}{\rm tr}
 ( \boldsymbol{H}_{\rm  eff} \boldsymbol{H}_{ \rm eff}^H ) ),
 \label{obj-tr}
\end{align}
where $(a)$ is due to Jensen's inequality and $(b)$ becomes
equality when $N_s= Q$. Therefore, alternatively,
\cite{NingChen20} proposed to maximize the bound of the spectral
efficiency:
\begin{align}
\max_{\boldsymbol{\Phi}}  \quad & {\rm tr} (   \boldsymbol{H}_{\rm
eff}
\boldsymbol{H}_{ \rm eff}^H )  \nonumber \\
 {\rm s.t.} \quad  & \boldsymbol{\Phi} = {\rm diag}(e^{j\phi_1},
    \cdots, e^{j \phi_M}).
\label{opt4}
\end{align}
Nevertheless, the objective of (\ref{opt4}) does not consider the
fairness among different singular values and thus results in a
large condition number of $\boldsymbol{H}_{\rm eff}$. A large
condition number indicates an unfavorable propagation condition
since we prefer the power of the channel to be uniformly
distributed over singular values to support multi-stream
transmission \cite{Marzetta16}. To address this issue, we will
develop a new approach to solve (\ref{opt3}). Our proposed method
obtains more balanced singular values of $\boldsymbol{\Sigma}_1$
which can help substantially improve the spectral efficiency, as
suggested by our empirical results.

\section{Proposed Passive Beamforming Design Method} \label{sec-LIS-BF}
Optimizing (\ref{opt3}) is much more challenging than directly
optimizing the Frobenius-norm of ${\boldsymbol{H}}_{\rm eff}$, as
the singular value can not be expressed by $\boldsymbol{H}_{\rm
eff}$ in an explicit way. To address this difficulty, we exploit
the structure of mmWave channels to gain insight into the SVD of
the effective channel. Recall (\ref{BS-IRS})--(\ref{IRS-UE}) and
assume $|\beta_1| \geq |\beta_2| \geq \ldots\geq | \beta_L|$ and
$|\alpha_1| \geq |\alpha_2| \geq \cdots \geq |\alpha_P|$, where
$\alpha_i \triangleq \sqrt{\frac{N_t M}{P}} \tilde \alpha_i$, and
$\beta_i \triangleq \sqrt{\frac{M N_r}{L}} \tilde \beta_i$. We
will justify the order of $\{\beta_l\}$ and $\{\alpha_p\}$ later.
The effective channel $\boldsymbol{H}_{\rm eff}=\boldsymbol{R}
\boldsymbol{\Phi} \boldsymbol{G} $ can be written as
\begin{align}
&\boldsymbol{R} \boldsymbol{\Phi} \boldsymbol{G} \nonumber \\
=& \bigg(\sum_{i=1}^L \beta_i \boldsymbol{a}_{R_{  \rm UE}}
(\gamma_i^r) \boldsymbol{a}_{T_{ \rm  LIS}}^H(\theta_i^t,
\eta_i^t)\bigg) \boldsymbol{\Phi} \nonumber\\
&\quad \times \bigg(\sum_{j=1}^P \alpha_j \boldsymbol{a}_{R_{\rm
LIS}}(\theta_j^r, \eta_j^r)
\boldsymbol{a}_{T_{  \rm BS}}^H(\gamma_j^t)\bigg)   \nonumber \\
= & \sum_{i=1}^L  \sum_{j=1}^P   \beta_i \alpha_j
\boldsymbol{a}_{R_{\rm UE}}(\gamma_i^r)
\underbrace{\boldsymbol{a}_{T_{  \rm LIS}}^H(\theta_i^t, \eta_i^t)
\boldsymbol{\Phi} \boldsymbol{a}_{R_{ \rm  LIS}}(\theta_j^r,
\eta_j^r)}_{d_{ij}}
\boldsymbol{a}_{T_{ \rm  BS}}^H(\gamma_j^t)   \nonumber \\
=& \sum_{i=1}^L  \sum_{j=1}^P   \beta_i  \alpha_j d_{ij}
\boldsymbol{a}_{R_{ \rm  UE}}(\gamma_i^r) \boldsymbol{a}_{T_{\rm BS}}^H(\gamma_j^t)   \nonumber \\
=& \boldsymbol{A}_{R_{ \rm  UE}} \boldsymbol{D}
\boldsymbol{A}_{T_{\rm   BS}}^H, \label{H-eff}
\end{align}
where ${\boldsymbol{D}}$ is an $L\times P$ matrix with its
$(i,j)$th entry given by ${\boldsymbol{D}(i,j)} = \beta_i \alpha_j
d_{ij}$, and $d_{ij}$ is defined as
\begin{align}
d_{ij} \triangleq& \boldsymbol{a}_{T_{\rm LIS}}^H(\theta_i^t,
\eta_i^t)
\boldsymbol{\Phi}  \boldsymbol{a}_{R_{\rm LIS}}(\theta_j^r, \eta_j^r) \nonumber \\
=&  \boldsymbol{v}^H  (\boldsymbol{a}_{T_{\rm
LIS}}^{\ast}(\theta_i^t, \eta_i^t)\circ
\boldsymbol{a}_{R_{\rm LIS}}(\theta_j^r, \eta_j^r))  \nonumber \\
= & \boldsymbol{v}^H \boldsymbol{p}^{ij} \label{dij}
\end{align}
with $\circ$ stands for the Hadamard (elementwise) product,
${\boldsymbol{v}\triangleq {\text{diag}}({\boldsymbol{\Phi}^H})}$,
and ${\boldsymbol{p}}^{ij} \triangleq (\boldsymbol{a}_{T_{\rm
LIS}}^{\ast}(\theta_i^t, \eta_i^t)\circ \boldsymbol{a}_{R_{\rm
LIS}}(\theta_j^r, \eta_j^r))$. Here $d_{ij}$ is referred to as the
\emph{passive beamforming gain} associated with the $(j,i)$th
BS-LIS-UE composite path which is composed of the $j$th path from
the BS to the LIS and the $i$th path from the LIS to the UE.

Also, $\boldsymbol{A}_{R_{ \rm UE}}$ and $\boldsymbol{A}_{T_{\rm
BS}}$ in (\ref{H-eff}) are respectively defined as
\begin{align}
\boldsymbol{A}_{R_{ \rm  UE}} \triangleq & [ \boldsymbol{a}_{R_{
\rm UE}}(\gamma_1^r) \phantom{0}\cdots\phantom{0}
\boldsymbol{a}_{R_{  \rm UE}}(\gamma_L^r)   ], \nonumber \\
\boldsymbol{A}_{T_{\rm   BS}} \triangleq & [\boldsymbol{a}_{T_{\rm
BS}}(\gamma_1^t)\phantom{0}\cdots\phantom{0} \boldsymbol{a}_{T_{
\rm BS}}(\gamma_P^t)].
\end{align}
For ULA with $N$ antennas, it can be easily verified that
\begin{align}
|{\boldsymbol{a}}^H(\gamma_i) {\boldsymbol{a}}(\gamma_j)|
\rightarrow 0, \quad N \rightarrow \infty
\end{align}
for any $\sin(\gamma_i) - \sin(\gamma_j) \neq 0$. This asymptotic
orthogonality also holds valid for other array geometries such as
UPA \cite{Chen13}. Therefore when $N_t$ and $N_r$ are sufficiently
large, $\boldsymbol{A}_{R_{\rm UE}}$ and $\boldsymbol{A}_{T_{\rm
BS}}$ can be considered as orthonormal matrices in which the
columns vectors form an orthonormal set (each column vector has
unit norm and is orthogonal to all the other column vectors). If
the phase shift vector $\boldsymbol{v}$ is properly devised such
that the off-diagonal elements of $\boldsymbol{D}$ are small
relative to entries on the main diagonal, then
$\boldsymbol{H}_{\rm eff}=\boldsymbol{A}_{R_{ \rm
UE}}\boldsymbol{D}\boldsymbol{A}_{T_{\rm BS}}^H$ can be treated as
an approximation of the truncated SVD of $\boldsymbol{H}_{\rm
eff}$, in which case the optimization problem \eqref{opt3} turns
into
\begin{align}
\max_{\boldsymbol{v}} \quad & \sum_{i=1}^{{  N_s}} \log_2\left(1+
\frac{\rho}{N_s \sigma^2}
|\boldsymbol{D}(i,i)|^2\right)  \nonumber \\
{\rm s.t.}  \quad &  \boldsymbol{D}(i,i) = \alpha_i\beta_i d_{ii}
, \quad  \forall i \in \{ 1, \cdots, N_s \}, \nonumber \\
& d_{ii}=\boldsymbol{v}^H \boldsymbol{p}^{ii}, \nonumber\\
&  |d_{ij}|=|\boldsymbol{v}^H \boldsymbol{p}^{ij}|<\tau, \quad \forall i\neq j, \nonumber\\
&\boldsymbol{v} =[e^{j\phi_1}\phantom{0}\cdots\phantom{0} e^{j
\phi_M}]^H, \label{opt5}
\end{align}
where $\tau$ is a small positive value and the constraint
$|d_{ij}|<\tau$ is to make sure that $\boldsymbol{D}$ is
approximately a non-square diagonal matrix such that
$\boldsymbol{H}_{\rm eff}=\boldsymbol{A}_{R_{ \rm
UE}}\boldsymbol{D}\boldsymbol{A}_{T_{\rm BS}}^H$ approximates a
truncated SVD of $\boldsymbol{H}_{\rm eff}$. One could question
that such a constraint (i.e. $|d_{ij}|<\tau,\forall i\neq j$) not
only complicates the problem, but also confines the solution space
of $\boldsymbol{v}$, and thus may prevent from achieving a higher
spectral efficiency. We will show that this constraint can be
ignored and this omission does not affect the validity of our
proposed solution. Specifically, we focus on solving
\begin{align}
\max_{\boldsymbol{v}} \quad & \sum_{i=1}^{{N_s}}\log_2\left(1+
\frac{\rho}{N_s \sigma^2}
|\boldsymbol{D}(i,i)|^2\right)  \nonumber \\
{\rm s.t.}  \quad &  \boldsymbol{D}(i,i) = \alpha_i\beta_i d_{ii}
, \quad  \forall i \in \{ 1, \cdots, N_s \} ,\nonumber \\
& d_{ii}=\boldsymbol{v}^H \boldsymbol{p}^{ii}, \nonumber\\
&\boldsymbol{v} =[e^{j\phi_1}\phantom{0}\cdots\phantom{0} e^{j
\phi_M}]^H, \label{opt6}
\end{align}
and we will show that for a sufficiently large $M$, the solution
to (\ref{opt6}) automatically ensures that off-diagonal entries of
$\boldsymbol{D}$ are small relative to entries on its main
diagonal. Thus it is expected that solving (\ref{opt6}) leads to
an effective solution of \eqref{opt3}.

%(which is the case for LIS consisting of a large number of passive
%reflecting elements)

\subsection{Manifold-Based Method for Passive Beamforming Design} \label{sec-man}
In this subsection, we develop a manifold-based method to solve
(\ref{opt6}). Note that other methods such as the semidefinite
relaxation (SDR) method, the block coordinate descent (BCD) method
\cite{ShiRazaviyayn11} and the penalty dual decomposition (PDD)
method \cite{ShiHong18} can also be employed to address
(\ref{opt6}). The reason we choose to use the manifold
optimization technique is that it achieves a good balance between
the computational complexity and the performance. Recalling that
$\boldsymbol{D}(i,i) = \alpha_i\beta_i \boldsymbol{v}^H
\boldsymbol{p}^{ii}$, the optimization (\ref{opt6}) can be
rewritten as
\begin{align}
 \max_{\boldsymbol{v}} \quad & \sum_{i=1}^{{  N_s}} \log_2(1+ a_i\boldsymbol{v}^H
 \boldsymbol{P}^{ii} \boldsymbol{v})  \nonumber \\
\text{s.t.} \quad & \boldsymbol{v}=
[e^{j\phi_1}\phantom{0}\cdots\phantom{0} e^{j \phi_M}]^H,
\label{opt-man2}
\end{align}
where $\boldsymbol{P}^{ii}\triangleq
\boldsymbol{p}^{ii}(\boldsymbol{p}^{ii})^H$, and $a_i\triangleq
\frac{\rho}{N_s \sigma^2} |\alpha_i\beta_i|^2$. The main
difficulty involved in solving (\ref{opt-man2}) is the non-convex
unit modulus constraint. We employ the manifold optimization
technique to address this difficulty. The search space of
\eqref{opt-man2} can be regarded as the product of $M$ complex
circles, i.e.,
\begin{align}
\underbrace{\mathcal{S} \times \mathcal{S} \cdots
\times\mathcal{S}  }_{M \; {\rm times}},
\end{align}
where $\mathcal{S} \triangleq \{ u \in \mathbb C: u^Hu = 1\}$ is
one complex circle\cite{AlhujailiMonga19}. The product of such $M$
circles is a submanifold of $\mathbb C^{M}$, known as the
\emph{complex circle manifold} (CCM) \cite{AbsilMahony09}, which
is defined as
\begin{align}
\mathcal{M} = \mathcal{S}^{M} \triangleq \{ \boldsymbol{u} \in
\mathbb C^{M}: |u_i |= 1, i =1, 2, \cdots,M\}.
\end{align}
More background on optimization on manifolds can be found in
\cite{AbsilMahony09}.

For a smooth real-valued objective function on some manifold, many
classical line-search algorithms such as the gradient descent
method can be applied with certain modifications
\cite{BoumalMihra14}. We start with some basic definitions. The
tangent space ${\mathcal{T}_{\boldsymbol{v}^k} \mathcal{M}}$ is
the set consisting of all tangent vectors to $\mathcal{M}$ at the
point $\boldsymbol{v}^k$ \cite{AbsilMahony09}. Here, the tangent
space of the complex circle manifold admits a closed-form
expression, i.e.,
\begin{align}
{\mathcal{T}_{\boldsymbol{v}^k} \mathcal{M}} = \{ \boldsymbol{z}
\in \mathbb C^{M}: \Re\{ \boldsymbol{z} \circ
(\boldsymbol{v}^k)^{\ast}\} = \boldsymbol{0}\},
\end{align}
where $\circ$ denotes the Hadamard product. Since the descent is
performed on the manifold, we need to find the direction of the
greatest decrease from the tangent space, which is known as the
negative \emph{Riemannian gradient}. For the complex circle
manifold, the Riemannian gradient of the objective function
$f(\boldsymbol{v}) \triangleq  -\sum_{i=1}^{{ N_s}} \log(1+ a_i
\boldsymbol{v}^H \boldsymbol{P}^{ii} \boldsymbol{v})$ at the point
$\boldsymbol{v}^k$ is a tangent vector $\nabla_{{\mathcal{M}}}
f(\boldsymbol{v}^k)$ given by \cite{AbsilMahony09}
\begin{align}
\nabla_{{\mathcal{M}}} f(\boldsymbol{v}^k) =& {\rm
Proj}_{\mathcal{T}_{\boldsymbol{v}^k}
\mathcal{M}} ( \nabla f( \boldsymbol{v}^k) ) \nonumber \\
=&   \nabla f( \boldsymbol{v}^k) - \Re \{ \nabla f(
\boldsymbol{v}^k) \circ (\boldsymbol{v}^k)^{\ast}  \}\circ
\boldsymbol{v}^k ,\label{Rie-grad}
\end{align}
where ${\rm Proj} (\cdot)$ is the orthogonal projection operator
and the Euclidean gradient $\nabla f( \boldsymbol{v}^k)$ is given
as
\begin{align}
\nabla f( \boldsymbol{v}^k) = - \sum_{i=1}^{N_s} \frac{1}{\ln
2}\frac{2a_i \boldsymbol{P}^{ii} \boldsymbol{v}^k}{1+ a_i
(\boldsymbol{v}^k)^H \boldsymbol{P}^{ii} \boldsymbol{v}^k}.
\label{Eu-grad}
\end{align}

After we obtain the Riemannian gradient, we can transplant many
line-search algorithms, such as the gradient descent method, from
the Euclidean spaces to the Riemannian manifolds
\cite{BoumalMihra14}. Specifically, we update $\boldsymbol{v}^{k}$
with a step $\varpi^k$
\begin{align}
\boldsymbol{\bar v}^{k} = \boldsymbol{v}^k - \varpi^k
\nabla_{{\mathcal{M}}} f(\boldsymbol{v}^k) ,\label{upd-vk}
\end{align}
where $\varpi^k$ is chosen as the Armijo step size
\cite{AbsilMahony09}. In general, $\boldsymbol{\bar v}^k$  does
not lie in the complex circle manifold, thus the \emph{retraction}
is needed to  map the updated point $\boldsymbol{\bar v}^{k}$ on
the tangent space $\mathcal{T}_{\boldsymbol{v}^k} \mathcal{M}$
onto the complex circle manifold $\mathcal{M}$ with a local
rigidity condition preserving the gradients at $\boldsymbol{v}^k$.
In this case, the retraction operator is defined as
\cite{AbsilMahony09}
\begin{align}
\boldsymbol{v}^{k+1} =  {\rm R}(\boldsymbol{\bar v}^k) \triangleq
\boldsymbol{\bar v}^k \circ \frac{1}{|\boldsymbol{\bar v}^k |}.
\label{upd-v}
\end{align}
The algorithm above is summarized in Algorithm \ref{AlgorithmP2},
which is guaranteed to converge to a critical point of
\eqref{opt-man2} \cite{AbsilMahony09}.

\begin{algorithm}[t!]
\caption{Proposed Algorithm for Solving {(\ref{opt-man2})
}}\label{AlgorithmP2}
\begin{algorithmic}[1]
\STATE  Initialize $\boldsymbol{v}^0 \in \mathcal{M}$ and give a
pre-defined threshold $\varepsilon$.
\REPEAT
\STATE Compute the
Euclidean gradient $\nabla f(\boldsymbol{v}^k)$ using
\eqref{Eu-grad};
\STATE Compute the Riemannian gradient
$\nabla_{{\mathcal{M}}} f(\boldsymbol{v}^k)$ with \eqref{Rie-grad};
\STATE Choose an Armijo step size $\varpi^k$ (\cite{AbsilMahony09},
see section 4.2.2);
\STATE Update $\boldsymbol{\bar v}^k =
\boldsymbol{v}^k - \varpi^k \nabla_{{\mathcal{M}}}
f(\boldsymbol{v}^k) $;
\STATE Update $\boldsymbol{v}^{k+1} =
{\rm R}(\boldsymbol{\bar v}^k)$ using \eqref{upd-v};
\UNTIL{the gap of
the objective function values between two iterations is smaller
than $\varepsilon$ }.
\STATE  Obtain the optimal solution
${\boldsymbol{v}}^{\star}$.
\end{algorithmic}
\end{algorithm}

\subsection{Discussions} \label{sec-dis}
Next, we discuss for a sufficiently large $M$, why the constraint
$|d_{i,j}|<\tau,\forall i\neq j$ in (\ref{opt5}) can be ignored
and the solution to (\ref{opt6}) yields small values of
$\{d_{ij},\forall i\neq j\}$.

Define $k\triangleq\frac{2\pi d}{\lambda}$, $ f(\theta, \eta)
\triangleq\cos(\eta)\sin(\theta)$, $g(\eta)\triangleq \sin(\eta)$,
and $\boldsymbol{\bar p}^{ij} \triangleq
\sqrt{M}\boldsymbol{p}^{ij}$. We have
\begin{align}
\boldsymbol{\bar p}^{ij} \triangleq
&\sqrt{M}(\boldsymbol{a}_{T_{\rm LIS}}^{\ast}(\theta_i^t,
\eta_i^t)\odot \boldsymbol{a}_{R_{\rm
LIS}}(\theta_j^r, \eta_j^r)) \nonumber \\
=& \frac{1} {\sqrt{M}} [1 \phantom{0}\ldots\phantom{0}
e^{jk(( m_1-1) ( \Delta f_{ij})  + (m_2-1)(\Delta g_{ij})}\phantom{0} \nonumber \\
& \quad \quad \quad  \ldots\phantom{0} e^{jk(( M_y-1) ( \Delta
f_{ij}) + (M_z-1)(\Delta g_{ij})} ]^T,
\end{align}
where $\Delta f_{ij} \triangleq
f(\theta_j^r,\eta_j^r)-f(\theta_i^t,\eta_i^t)$ and $\Delta g_{ij}
\triangleq g(\eta_j^r)-g(\eta_i^t)$. Therefore, we can calculate
$|(\boldsymbol{\bar p}^{ij} )^H \boldsymbol{ \bar p}^{mn}|$ as
\begin{align}
& |(\boldsymbol{\bar p}^{ij} )^H \boldsymbol{ \bar p}^{mn}|  \nonumber \\
= &\frac{1} {M} \left | \sum_{m_1 = 1}^{M_y} \sum_{ m_2 = 1}^{M_z}
e^{jk(( m_1-1)
( \Delta F_{ijmn})  + (m_2-1)( \Delta G_{ijmn})}  \right |\nonumber \\
=&  \frac{1} {M} \left |\frac{ \sin \left(  \frac{1}{2} kM_y
\Delta F_{ijmn}\right)}{ \sin \left(  \frac{1}{2} k \Delta
F_{ijmn}  \right)} \frac{ \sin \left(  \frac{1}{2} kM_z \Delta
G_{ijmn}\right)}{ \sin \left(  \frac{1}{2} k \Delta G_{ijmn}
\right)} \right |,
\end{align}
where $ \Delta F_{ijmn}\triangleq \Delta f_{mn} - \Delta f_{ij}$
and $ \Delta G_{ijmn}\triangleq\Delta g_{mn} -\Delta g_{ij}$. Due
to the random distribution of AoA and AoD parameters, we have
$\Delta F_{ijmn} \neq 0$ or $\Delta G_{ijmn} \neq 0$ for any
$\{m,n\}\neq\{i,j\}$ with probability one. Thus we have
\begin{align}
\lim_{M \rightarrow \infty} \mid (\boldsymbol{\bar p}^{ij} )^H
\boldsymbol{ \bar p}^{mn}\mid  =0,  \forall m \neq i \parallel n
\neq j,
\end{align}
where $\parallel$ denotes the OR logic operation which means
either $m\neq i$ or $n\neq j$ is true. Also, it is easy to verify
that $ (\boldsymbol{\bar p}^{ij} )^H \boldsymbol{ \bar p}^{ij} =
1$.

Define $\mathcal{A} \triangleq  \{\boldsymbol{\bar p}^{ii}: i=1,
\ldots, N_s \}$ and $\mathcal{B}\triangleq \{\boldsymbol{\bar
p}^{mn}: \{m,n\}\in\mathcal{E}\}$, where
$\mathcal{E}\triangleq\mathcal{U}-\mathcal{I}$, in which
\begin{align}
\mathcal{U}\triangleq&\{\{m,n\}:m\in\{1,\ldots,L\}, n\in
\{1,\ldots,P\}\} \\
\mathcal{I}\triangleq&\{\{i,i\}: i\in\{1,\ldots,N_s\}\}
\end{align}
Let $\mathcal{C}\triangleq \{ \boldsymbol{c}_u\}_{u=1}^{M-LP}$
denote an orthonormal set whose vectors are orthogonal to those
vectors in $\mathcal{A} \cup \mathcal{B}$. Clearly, the set of
vectors in $\mathcal{D} \triangleq\mathcal{A} \cup \mathcal{B}
\cup \mathcal{C}$ form an orthonormal basis of $\mathbb C^{M}$.
Hence, any vector in $\mathbb C^{M}$ can be expressed as a sum of
the basis vectors scaled
\begin{align}
\boldsymbol{v} = \sum_{i=1}^{N_s} k_{ii} \boldsymbol{\bar
p}^{ii}+\sum_{\{m,n\}\in\mathcal{E}} k_{mn} \boldsymbol{\bar
p}^{mn} +\sum_{u=1}^{M-LP} k_{u} \boldsymbol{c}_{u}. \label{eqn4}
\end{align}
It should be noted that
\begin{align}
M= \boldsymbol{v}^H \boldsymbol{v} =&  \sum_{i=1} ^{N_s} | k_{ii}
| ^2  + \sum_{\{m,n\}\in\mathcal{E}} |k_{mn}|^2  +
\sum_{u=1}^{M-LP} |k_{u} |^2.
\end{align}
Substituting (\ref{eqn4}) into the objective function of
\eqref{opt-man2}, we have
\begin{align}
&   \sum_{i=1}^{N_s} \log_2 (1+ a_i| \boldsymbol{v}^H \boldsymbol{p}^{ii}|^2)  \nonumber \\
   = & \sum_{i=1}^{N_s} \log_2 \bigg (1+
   a_i  \bigg| \bigg( \sum_{i=1}^{N_s} k_{ii} \boldsymbol{\bar p}^{ii}+\sum_{\{m,n\}\in\mathcal{E}}
   k_{mn} \boldsymbol{\bar p}^{mn}  \nonumber \\
  & + \sum_{u=1}^{M-LP} k_{u} \boldsymbol{c}_{u}\bigg)^H \boldsymbol{p}^{ii} \bigg|^2 \bigg) \nonumber \\
   =& \sum_{i=1}^{N_s} \log_2\left(1 + \frac{a_i}{M}  | k_{ii}|^2 \right).
\end{align}
Therefore \eqref{opt-man2} can be re-expressed as
\begin{align}
 \max_{\{k_{ii},k_{mn},k_u\}} \quad & \sum_{i=1}^{N_s} \log_2(1 + \frac{a_i}{M}| k_{ii}|^2 )  \nonumber \\
\text{s.t.} \quad   &
 \sum_{i=1} ^{N_s} |k_{ii}|^2  \leq M, \nonumber\\
  &
\boldsymbol{v}=[e^{j\phi_1}\phantom{0}\cdots\phantom{0} e^{j
\phi_M}]^H, \nonumber\\
& \boldsymbol{v} = \sum_{i=1}^{N_s} k_{ii} \boldsymbol{\bar
p}^{ii}+\sum_{\{m,n\}\in\mathcal{E}} k_{mn} \boldsymbol{\bar
p}^{mn} +\sum_{u=1}^{M-LP} k_{u} \boldsymbol{c}_{u}.
 \label{opt-k}
\end{align}
If no unit modulus constraint is placed on the phase shift vector
$\boldsymbol{v}$, then it is clear that the objective function
value of \eqref{opt-k} is maximized when $\sum_{i=1}^{N_s}
|k_{ii}|^2=M$, which means that $k_{mn}=0,\forall
\{m,n\}\in\mathcal{E}$ and $k_u=0$. As a result, we have
\begin{align}
d_{mn}=(\boldsymbol{v}^{\star})^H\boldsymbol{p}^{mn}=\left(\sum_{i=1}^{N_s}
k_{ii} \boldsymbol{\bar p}^{ii}\right)^H\boldsymbol{p}^{mn}=0
\quad \forall \{m,n\}\in\mathcal{E}
\end{align}
Hence the solution to (\ref{opt-k}) or (\ref{opt-man2}),
$\boldsymbol{v}^{\star}$, yields exactly zero off-diagonal entries
of $\boldsymbol{D}$. With the unit modulus constraint
$\boldsymbol{v}=[e^{j\phi_1}\phantom{0}\cdots\phantom{0} e^{j
\phi_M}]^H$, $\sum_{i=1}^{N_s} |k_{ii}|^2$ will not be exactly equal
to $M$ but the solution will force $\sum_{i=1}^{N_s} |k_{ii}|^2$
to be close to $M$ as much as possible in order to obtain a
maximum objective function value. As a result, the values of
$\{|k_{mn}|\},\forall \{m,n\}\in\mathcal{E}$ are small relative to
$\{|k_{ii}|,i=1,\ldots,N_s\}$. Note that we have
\begin{align}
&d_{pq}=(\boldsymbol{v}^{\star})^H\boldsymbol{p}^{pq}\nonumber\\
=&\left(\sum_{i=1}^{N_s} k_{ii} \boldsymbol{\bar
p}^{ii}+\sum_{\{m,n\}\in\mathcal{E}} k_{mn} \boldsymbol{\bar
p}^{mn} +\sum_{u=1}^{M-LP} k_{u}
\boldsymbol{c}_{u}\right)^H\boldsymbol{p}^{pq} \nonumber\\
=&\begin{cases}
\frac{k_{pq}}{\sqrt{M}}    & q\neq p  \\
\frac{k_{pp}}{\sqrt{M}}  & q=p
\end{cases}
\end{align}
Therefore the optimal solution to (\ref{opt-man2}) yields an
approximately diagonal matrix $\boldsymbol{D}$ whose off-diagonal
entries $\{d_{ij},i\neq j\}$ are small relative to its diagonal
entries $\{d_{ii},i=1,\ldots,N_s\}$.

\subsection{Ordering of $\{\beta_i\}$ and $\{\alpha_i\}$}
Note that we assumed a decreasing order of the path gains
$\{\beta_i\}$ and $\{\alpha_i\}$ earlier in this paper, i.e.
$|\beta_1| \geq |\beta_2| \geq \ldots\geq | \beta_L|$ and
$|\alpha_1| \geq |\alpha_2| \geq \cdots \geq |\alpha_P|$.
Nevertheless, from (\ref{opt6}), we can see that different orders
of $\{\beta_i\}$ and $\{\alpha_i\}$ may lead to different
solutions and thus different performance, and it is unclear
whether arranging the path gains $\{\beta_i\}$ and $\{\alpha_i\}$
in a decreasing order is a good choice. To gain an insight into
this problem, let us examine the optimization (\ref{opt-k}) which
is a variant of (\ref{opt6}). As discussed above, if we ignore
the unit-modulus constraint placed on $\boldsymbol{v}$,
(\ref{opt-k}) can be simplified as a conventional power allocation
problem:
\begin{align}
 \max_{\{\tilde{k}_{ii}\}} \quad & \sum_{i=1}^{N_s} \log_2(1 + a_i |\tilde{k}_{ii}|^2 )  \nonumber \\
\text{s.t.} \quad   &
 \sum_{i=1} ^{N_s} |\tilde{k}_{ii}|^2  \leq 1,
 \label{opt7}
\end{align}
where $\tilde{k}_{ii} \triangleq  \frac{k_{ii}}{\sqrt{M}}$. Recall
that $a_i=\frac{\rho}{N_s\sigma^2}|\alpha_i\beta_i|^2$, where
$\alpha_i=\sqrt{\frac{N_{t}M}{P}}\tilde{\alpha}_i$, and
$\beta_i=\sqrt{\frac{N_{r}M}{L}}\tilde{\beta}_i$. Thanks to the
large dimensions of $N_t$, $N_r$, and $M$, it can be expected that
the effective signal-to-noise ratios (SNRs) $\{a_i\}$ have large
values even the nominal SNR $\frac{\rho}{\sigma^2}$ is small.
It is well-known that in the high SNR regime, it is approximately
optimal to equally allocate power to different data streams.
Therefore the question is how to arrange the orders of
$\{\beta_i\}$ and $\{\alpha_i\}$ such that $\sum_{i=1}^{N_s}
\log_2(1+ a_i/N_s)$ can be maximized.

Let $\{\tau_l\}_{l=1}^L$ be a permutation of the set
$\{1,\ldots,L\}$ and $\{\pi_p\}_{p=1}^P$ a permutation of the set
$\{1,\ldots,P\}$. We can write
\begin{align}
S\triangleq&\sum_{i=1}^{N_s} \log_2(1+ a_i/N_s)=\sum_{k=1}^{N_s}
\log_2
(1+ e | \alpha_{\pi_k}|^2 | \beta_{\tau_k}|^2) \nonumber \\
=& \sum_{k=1}^{N_s} \log_2 ( | \alpha_{\pi_k}|^2) +
\sum_{k=1}^{N_s} \log_2 \left( \frac{1}{ |\alpha_{\pi_k} |^2} + e |\beta_{\tau_k}|^2\right) \nonumber \\
\triangleq & S_1 + S_2.
\end{align}
where $e \triangleq \frac{\rho}{N_s^2 \sigma^2}$.
We see that $S_1$ is a constant once
$\{\alpha_{\pi_k}\}_{k=1}^{N_s}$ is determined. Hence, we focus
on the value of $S_2$. Note that
\begin{align}
2^{S_2} = \prod_{k=1}^{N_s} \left( \frac{1}{ |\alpha_{\pi_k}|^2} +
e | \beta_{\tau_k}|^2\right).
\end{align}
Without loss of generality, we assume
\begin{align}
\frac{1}{|\alpha_{\pi_1} |^2} \leq \frac{1}{|\alpha_{\pi_2}|^2}
\leq \cdots \leq \frac{1}{|\alpha_{\pi_{N_s}}|^2}.
\end{align}
Clearly, according to the dual rearrangement inequality
\cite{Oppenheim54}, $S_2$ attains its maximum if and only if
\begin{align}
|\beta_{\tau_1}|^2 \geq |\beta_{\tau_2}|^2 \geq \cdots \geq
|\beta_{\tau_{N_s}}|^2.
\end{align}
This means that similar to $\{\alpha_{\pi_k}\}_{k=1}^{N_s}$, $\{|
\beta_{\tau_k}|\}_{k=1}^{N_s}$ should also be arranged in a
decreasing order. This explains why we arrange the path gains of
$\{\beta_i\}$ and $\{\alpha_i\}$ in a descending order.

\section{Hybrid Precoding/Combining Design} \label{sec-hyb-man}
After obtaining the passive beamforming matrix
$\boldsymbol{\Phi}^{\star}$, we conduct the SVD of
$\boldsymbol{H}^{\star}$, i.e. $\boldsymbol{H}^{\star} =
\boldsymbol{R}\boldsymbol{\Phi}^{\star}\boldsymbol{G} =
\boldsymbol{U}^{\star} \boldsymbol{\Sigma}^{\star}
(\boldsymbol{V}^{\star})^H$ and the optimal precoder/combiner can
be obtained via \eqref{digi-opt}. As discussed earlier in this
paper, we search for an analog precoding (combining) matrix
$\boldsymbol{F}_{\text{RF}}$ ($\boldsymbol{W}_{\text{RF}}$) and a
baseband precoding (combining) matrix $\boldsymbol{F}_{\text{BB}}$
($\boldsymbol{W}_{\text{BB}}$) to approximate the optimal precoder
(combiner) $\boldsymbol{F}_{\text{opt}}$
($\boldsymbol{W}_{\text{opt}}$). In the following, we focus our
discussion on the hybrid precoding design as the extension to the
hybrid combining design is similar. Such a problem can be
formulated as \cite{Kasai18}
\begin{align}
\min_{ \boldsymbol{w}\in {\mathcal{M}_2}, \boldsymbol{F}_{\rm BB}}
\| \boldsymbol{F}_{\rm opt}-\boldsymbol F_{ \rm  RF} \boldsymbol
F_{\rm  BB} \|_F^2, \label{opt-hyb-2}
\end{align}
where $\boldsymbol{w}\triangleq{\rm vec}(\boldsymbol{F}_{\rm RF})$
and the search space is a complex circle manifold defined as
$\mathcal{M}_2 \triangleq \mathcal{S}^{N_t R_t}$. Specifically, we
adopt the fast manifold-based optimization method in
\cite{Kasai18} to solve (\ref{opt-hyb-2}), where we optimize
$\boldsymbol{F}_{\text{RF}}$ and $\boldsymbol{F}_{\text{BB}}$ in
an alternating manner. The detailed procedure can be found in
\cite{Kasai18} and thus omitted here. It should be noted that the
method is guaranteed to converge to a critical point
\cite{Kasai18,AbsilMahony09} and has a computational complexity at
the order of $\mathcal{O}(N_tR_tN_sL_2)$, where $L_2$ is the
number of iterations required to converge.

\renewcommand{\algorithmicrequire}{ \textbf{Input:}} %Use Input in the format of Algorithm
\renewcommand{\algorithmicensure}{ \textbf{Output:}}
\begin{algorithm}[t!]
\caption{Proposed Algorithm for Joint Transceiver and LIS
Design}\label{AlgorithmP3}
\begin{algorithmic}[1]
\REQUIRE { $\boldsymbol{R}$, $\boldsymbol{G}$, $N_s$, $R_t$,
$R_r$, $\rho$, $\sigma^2$};
\\   \textit{First step}
\STATE  Optimizing $\boldsymbol{\Phi}^{\star}$ by Algorithm
\ref{AlgorithmP2} developed in Subsection \ref{sec-man};
\\  \textit{Second step}
\STATE Compute the optimal precoding and combining matrices, i.e.
$\boldsymbol{F}_{\rm opt}$ and $\boldsymbol{W}_{\rm opt}$, using
\eqref{digi-opt} \STATE Obtain hybrid precoding and combining
matrices, i.e. $\boldsymbol{F}_{\rm BB}$, $\boldsymbol{F}_{\rm
RF}$  $\boldsymbol{W}_{\rm BB}$ and $\boldsymbol{W}_{\rm RF}$, via
the manifold-based method discussed in Section \ref{sec-hyb-man};
\ENSURE {$\boldsymbol{\Phi}^{\star}$, $\boldsymbol{F}_{\rm BB}$,
$\boldsymbol{F}_{\rm RF}$, $\boldsymbol{W}_{\rm BB}$, and
$\boldsymbol{W}_{\rm RF}$.}
\end{algorithmic}
\end{algorithm}

%As a result, it can only be applied to mmWave channel scenarios
%where both $\boldsymbol{G}$ and $\boldsymbol{R}$ are characterized
%by the Saleh-Valenzuela (SV) channel model.

\section{Summary and Discussions} \label{sec-comp}
First of all, we would like to clarify that different from other
state-of-the-art methods \cite{ZhangZhang19,NingChen20}, our
proposed method is designed specifically for mmWave systems. It
relies on the sparse scattering structure of mmWave channels to
obtain a good approximation of the SVD of the effective channel.
On the other hand, as will be shown in the subsequent analysis and
simulation results, utilizing the inherent sparse structure of
mmWave channels enables us to achieve a better tradeoff between
the spectral efficiency performance and the computational
complexity.

\subsection{Computational Complexity Analysis}
For clarity, the proposed algorithm is summarized in Algorithm
\ref{AlgorithmP3}. Specifically, our proposed algorithm consists
of two steps. The first step aims to obtain LIS's reflection
coefficients via solving the passive beamforming optimization
problem \eqref{opt-man2}. A manifold-based method, i.e. Algorithm
\ref{AlgorithmP2}, is developed. The dominant operation for
Algorithm \ref{AlgorithmP2} at each iteration is to calculate the
Euclidean gradient \eqref{Eu-grad}, which has a computational
complexity at the order of $\mathcal{O}(M)$. Note that
$\boldsymbol{P}^{ii} = \boldsymbol{p}^{ii}
(\boldsymbol{p}^{ii})^H$, and thus calculating
$(\boldsymbol{p}^{ii})^H \boldsymbol{v}$ is enough to obtain
\eqref{Eu-grad}. Therefore the first step involves a computational
complexity of $\mathcal{O}(M L_1)$, where $L_1$ denotes the number
of iterations required to converge. The second step solves a
hybrid beamforming problem which involves calculating the optimal
precoder/combiner and searching for an analog precoding
(combining) matrix and a baseband precoding (combining) matrix to
approximate the optimal precoder (combiner). The optimal
precoder/combiner can be obtained by performing the SVD of the
effective channel, whose complexity is at the order of
$\mathcal{O}(N_rN_t\min(N_r,N_t))$. The proposed manifold-based
algorithm for finding the analog/diginal precoding (combining)
matrices, as discussed in the previous section, has a complexity
of $\mathcal{O}(N_r R_r N_s L_2+N_t R_t N_s L_2)$, where $L_2$
denotes the number of iterations required to converge. Thus, the
overall computational complexity of our proposed algorithm is at
the order of $\mathcal{O}(ML_1+N_rN_t\min(N_r,N_t) +  N_r R_r N_s
L_2+N_t R_t N_s L_2)$.

Note that there is no alternating process between these two steps.
We only need to execute each step once for our proposed algorithm.
Since each step of our proposed algorithm is ensured to converge,
the convergence of our proposed algorithm is guaranteed.

\subsection{Computational Complexity Comparison}
In this subsection, we compare our proposed method with some
state-of-the-art methods developed for spectral efficiency
optimization for IRS-aided MIMO systems. Note that these methods
are not specially designed to mmWave systems with hybrid
precoding/combining structures. To apply them to our problem, a
similar two-step procedure is required. Let us first consider the
alternating optimization (AO)-based algorithm \cite{ZhangZhang19}.
In the first step, the AO-based algorithm \cite{ZhangZhang19} is
used to optimize the LIS's reflection coefficients
$\boldsymbol{\Phi}$ and the transmit covariance matrix
$\boldsymbol{Q}$. The first step involves a computational
complexity at the order of
$\mathcal{O}(N_rN_t(M+\min(N_r,N_t))L+((3N_r^3 +2N_r^2N_t+N_t^2)M+
N_rN_t\min(N_r,N_t))I)$, where $L$ is the number of realizations
for initialization and $I$ denotes the number of outer iterations
\cite{ZhangZhang19}. Note that the first step requires to
calculate the SVD of the effective channel. Hence the optimal
precoder/combiner can be directly obtained without involving
additional computational complexity. In the second step, similar
to our work, we resort to the manifold-based algorithm to find the
hybrid precoder and combiner to approximate the optimal
precoder/combiner, which involves a same computational complexity
at the order of $\mathcal{O}(N_rR_rN_sL_2 + N_t R_t N_s L_2)$.
Therefore the overall computational complexity of the AO-based
algorithm \cite{ZhangZhang19} is at the order of
$\mathcal{O}(N_rN_t(M+\min (N_r,N_t))L + ((
3N_r^3+2N_r^2N_t+N_t^2)M + N_rN_t\min (N_r,N_t))I + N_rR_rN_sL_2 +
N_tR_tN_sL_2)$. Note that although both the AO-based algorithm
\cite{ZhangZhang19} and our proposed algorithm have a
computational complexity scaling linearly with $M$, our proposed
algorithm achieves a substantial complexity reduction as compared
with \cite{ZhangZhang19}. To see this, suppose $N_t\geq N_r$ and
$N_r\gg \min\{R_r,R_t\}\geq N_s$, which is a reasonable setup for
practical systems. In this case, the complexity of our proposed
method is dominated by $\mathcal{O}(ML_1+N_r^2N_t)$. As a
comparison, the computational complexity of the AO-based algorithm
\cite{ZhangZhang19} is dominated by $\mathcal{O}(2N_r^2N_t M I)$.
Considering the fact that $L_1$\footnote{Simulation results show
that our proposed manifold-based algorithm converges within only a
few, say, ten iterations.} is much smaller than $2N_r^{2}N_{t}I$,
our proposed method has a much lower computational complexity than
the AO-based method \cite{ZhangZhang19}.

% The complexity of other existing methods can be similarly
% analyzed.

% In Table \ref{table-comp}, we summarize the
% computational complexity of our proposed method and other
% competing algorithms including the AO-based method
% \cite{ZhangZhang19}, the sum-path-gain maximization (SPGM) method
% \cite{NingChen20} and the weighted minimum mean square error
% (WMMSE) method \cite{PanRen19}. Here $I_2$ ($T$) denotes the outer
% (inner) iterations of the WMMSE-based algorithm \cite{PanRen19},
% and $L_3$ denotes the number of iterations of SPGM-based algorithm
% in \cite{NingChen20}. We see that the computational complexity of
% both the WMMSE \cite{PanRen19} and the SPGM \cite{NingChen20}
% scales cubically with the number of reflecting elements $M$.

The complexity of other existing methods can be similarly
analyzed. Specifically, the overall computational complexity of
the weighted minimum mean square error (WMMSE)-based method in
\cite{PanRen19} is at the order of $ \mathcal{O}( (N_r^3+N_s^3+
{\max (N_t^3, N_t^2N_r)} + (M^3+ TM^2) )I_2 + N_rR_rN_sL_2 +
N_tR_tN_sL_2)$, where $I_2$ ($T$) denotes the outer (inner)
iterations. On the other hand, the overall computational
complexity of the sum-path-gain maximization (SPGM)-based method
in \cite{NingChen20} is at the order of $\mathcal{O}( M^3L_3+
N_rN_t\min(N_r,N_t) +  N_r R_r N_s L_2+N_t R_t N_s L_2)$, where
$L_3$ denotes the number of iterations. Assuming $M\geq
N_t>N_r>N_s$, the dominant computational complexity of respective
algorithms is summarized in Table \ref{table-comp}. We see that
the computational complexity of both \cite{PanRen19} and
\cite{NingChen20} scales cubically with the number of reflecting
elements $M$.

%the WMMSE-based method requires much more computational resources
%owing to its slow convergence, while the SPGM-based method
%consumes much smaller resources thanks to its fast convergence.

\begin{table}[h!]
\centering \caption{Computational complexity comparison}
\begin{tabular}{c|c}
\hline
Algorithm & Dominant computational complexity \\
\hline
T-SVD-BF (proposed) & $\mathcal{O}(ML_1+N_r^2N_t)$.\\
AO-based algorithm in \cite{ZhangZhang19} & $\mathcal{O}(2N_r^2N_t M I)$\\
WMMSE-based algorithm in \cite{PanRen19} &
$\mathcal{O}((M^3+TM^2)I_2)$\\
SPGM-based algorithm in \cite{NingChen20} & $\mathcal{O}(M^3L_3)$\\
\hline
\end{tabular}
\label{table-comp}
\end{table}

\begin{figure}[!t]
    \centering
    {\includegraphics[width=3.5in]{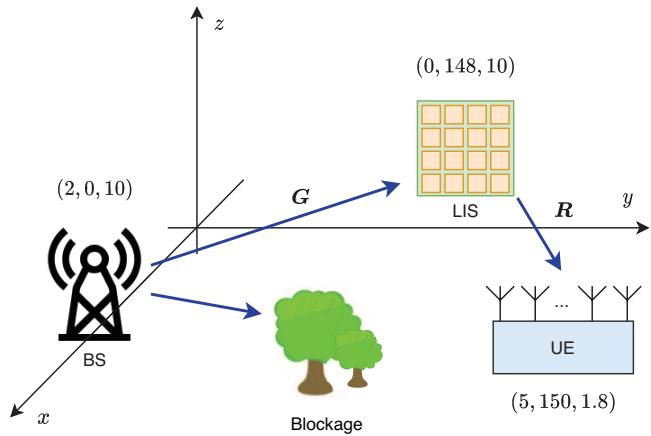}}
    \caption{Simulation setup.} \label{fig_Simusetup}
\end{figure}

\section{Simulation Results} \label{sec-simu}
In this section, we evaluate the performance of our proposed joint
transceiver and LIS design method. The proposed method is based on
the approximation of the truncated SVD of the effective (cascade)
channel, and thus is referred to as Truncated-SVD-based
beamforming (T-SVD-BF). In our simulations, we adopt a
three-dimensional setup as shown in Fig. \ref{fig_Simusetup}. The
coordinates of the BS, the LIS and the UE are respectively given
as $(x_{\rm BS},0,z_{\rm BS})$, $(0,y_{\rm LIS}, z_{\rm LIS})$,
and $(x_{\rm UE}, y_{\rm UE},z_{\rm UE})$, where we set $x_{\rm
BS} = 2$m, $z_{\rm BS} = 10$m, ${y}_{\rm LIS} = 148$m, $z_{\rm
LIS} = 10$m, $x_{\rm UE} = 5$m, $y_{\rm UE} = 150$m, and $z_{\rm
UE} = 1.8$m. The distance between the BS (LIS) and the LIS (UE)
can be easily calculated and given as $148$m ($9.8$m).

Slightly different from our previous definition, the effective
channel is modeled as follows by considering the transmitter and
receiver antenna gains \cite{NingChen19}
\begin{align}
\boldsymbol{H}_{\rm eff} = G_t G_r \boldsymbol{R}\boldsymbol{\Phi}
\boldsymbol{G}
\end{align}
where $G_t$ and $G_r$ represent the transmitter and receiver
antenna gains, respectively. The BS-LIS channel $\boldsymbol{G}$
and the LIS-UE channel $\boldsymbol{R}$ are modeled as follows
\begin{align}
\boldsymbol{G} = &\sqrt{ \frac{N_tM}{P}}  \bigg(
\tilde{\alpha}_1\boldsymbol{a}_{R_{   \rm  LIS}}(\theta_1^r,
\eta_1^r) \boldsymbol{a}_{T_{  \rm   BS}}^H(\gamma_1^t) \nonumber
\\ & + \sum_{i=2}^{P} \tilde\alpha_i \boldsymbol{a}_{R_{   \rm
LIS}}(\theta_i^r, \eta_i^r) \boldsymbol{a}_{T_{  \rm
BS}}^H(\gamma_i^t)  \bigg), \label{BS-IRS-Rician}
\\
\boldsymbol{R} = & \sqrt{ \frac{MN_r}{L}}  \bigg( \tilde \beta_1
\boldsymbol{a}_{R_{  \rm   UE}} (\gamma_1^r) \boldsymbol{a}_{T_{
\rm   LIS}}^H(\theta_1^t, \eta_1^t) \nonumber \\& + \sum_{i=2}^{L}
\tilde \beta_i \boldsymbol{a}_{R_{  \rm UE}} (\gamma_i^r)
\boldsymbol{a}_{T_{ \rm   LIS}}^H(\theta_i^t, \eta_i^t) \bigg),
\label{IRS-UE-Rician}
\end{align}
where $\tilde{\alpha}_1 (\tilde{\beta}_1) \sim {\cal
CN}(0,10^{-0.1 \kappa}) $ denotes the complex gain of the LOS
path, $\kappa$ is the path loss given by \cite{AkdenizLiu14}
\begin{align}
\kappa= a+10b \log_{10}(\tilde{d}) + \xi
\end{align}
in which $\tilde{d}$ denotes the distance between the transmitter
and the receiver, and $\xi\sim \mathcal{N}(0,\sigma_{\xi}^2)$. The
values of $a$, $b$ $\sigma_{\xi}$ are set to be $a=61.4$, $b=2$,
and $\sigma_{\xi}=5.8$dB as suggested by LOS real-world channel
measurements \cite{AkdenizLiu14}, $\tilde{\alpha}_i
(\tilde{\beta_i}) \sim \mathcal{CN} (0, 10^{-0.1(\kappa+\mu)})$
stands for the complex gain of the associated NLOS path, and $\mu$
is the Rician factor \cite{YingGao20,Samimi16}. Unless specified
otherwise, we assume that the LIS employs a UPA with $M=M_y \times
M_z= 16\times 16$, the BS and the UE adopt ULAs with $N_t = N_r =
64$. Other parameters are set as follows: $R_t= R_r= 6$, $N_s=4$,
$L=P=7$, $\mu = 10$, $G_t = 24.5$dBi, $G_r = 0$dBi. The carrier
frequency is set to $28$GHz, the bandwidth is set to $251.1886$MHz
and thus the noise power is $\sigma^2 =-174+10 \log_{10}B =
-90$dBm.

We compare our proposed method with the following three
state-of-the-art algorithms, namely, the sum-path-gain
maximization (SPGM) method which aims to maximize the
Frobenius-norm of the effective channel
$\boldsymbol{H}_{\text{eff}}$ \cite{NingChen20}, the AO-based
algorithm \cite{ZhangZhang19}, and the WMMSE-based algorithm
\cite{PanRen19}. Note that the WMMSE-based method is designed for
multicell-multiuser scenarios. But it can easily adapted to the
point-to-point scenario considered in this paper.

\begin{figure}[!t]
    \centering
    {\includegraphics[width=3.5in]{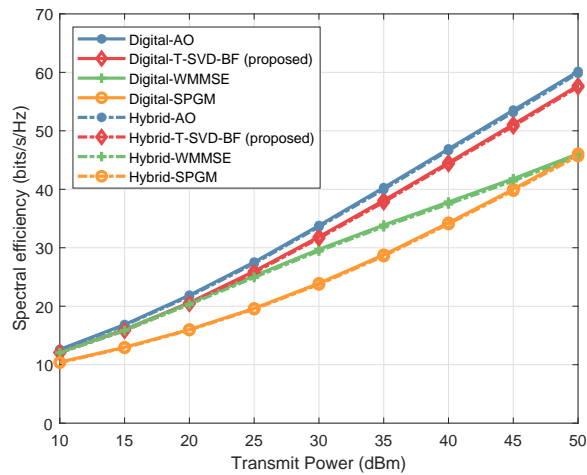}}
    \caption{Spectral efficiency versus transmit power $\rho$.} \label{figSEvSNR}
\end{figure}

\begin{figure}[!t]
    \centering
    {\includegraphics[width=3.5in]{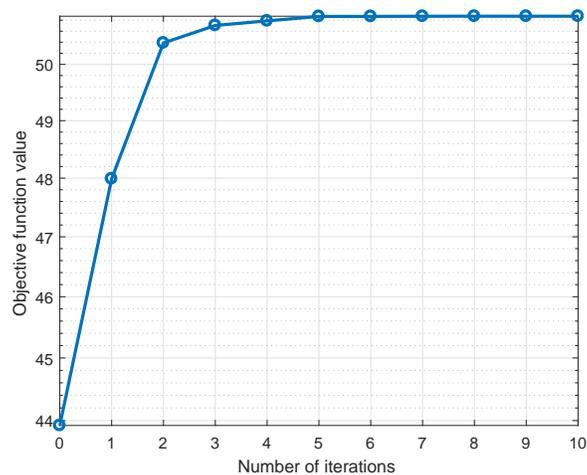}}
    \caption{Convergence behavior of the proposed algorithm for solving \eqref{opt-man2} } \label{fig_converge}
\end{figure}

\begin{figure*}[!t]
 \subfigure[Spectral efficiency versus the number of LIS elements $M$.]
 {\includegraphics[width=3.5in]{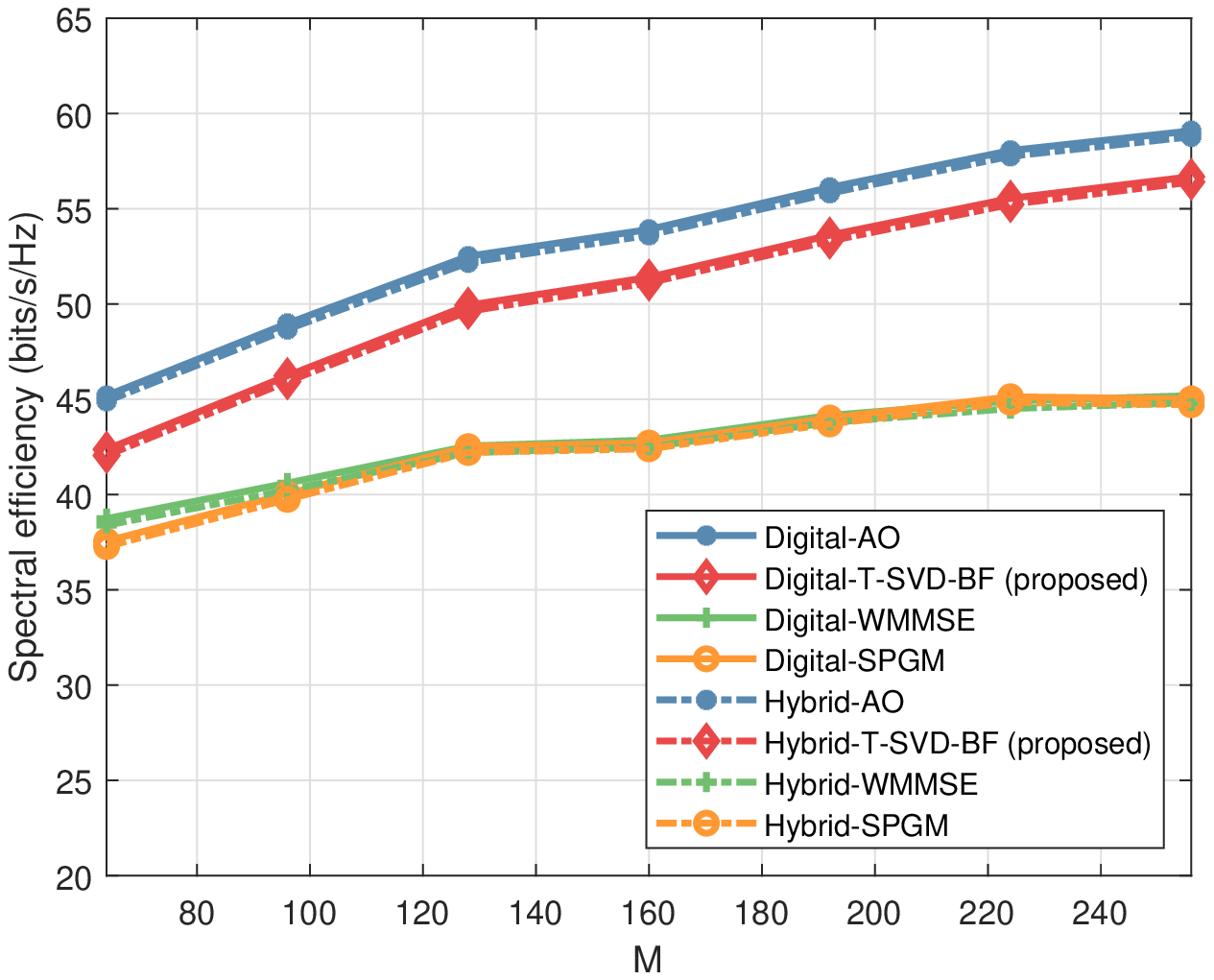}} \hfil
\subfigure[Average run time versus the number of LIS elements $M$.]
{\includegraphics[width=3.5in]{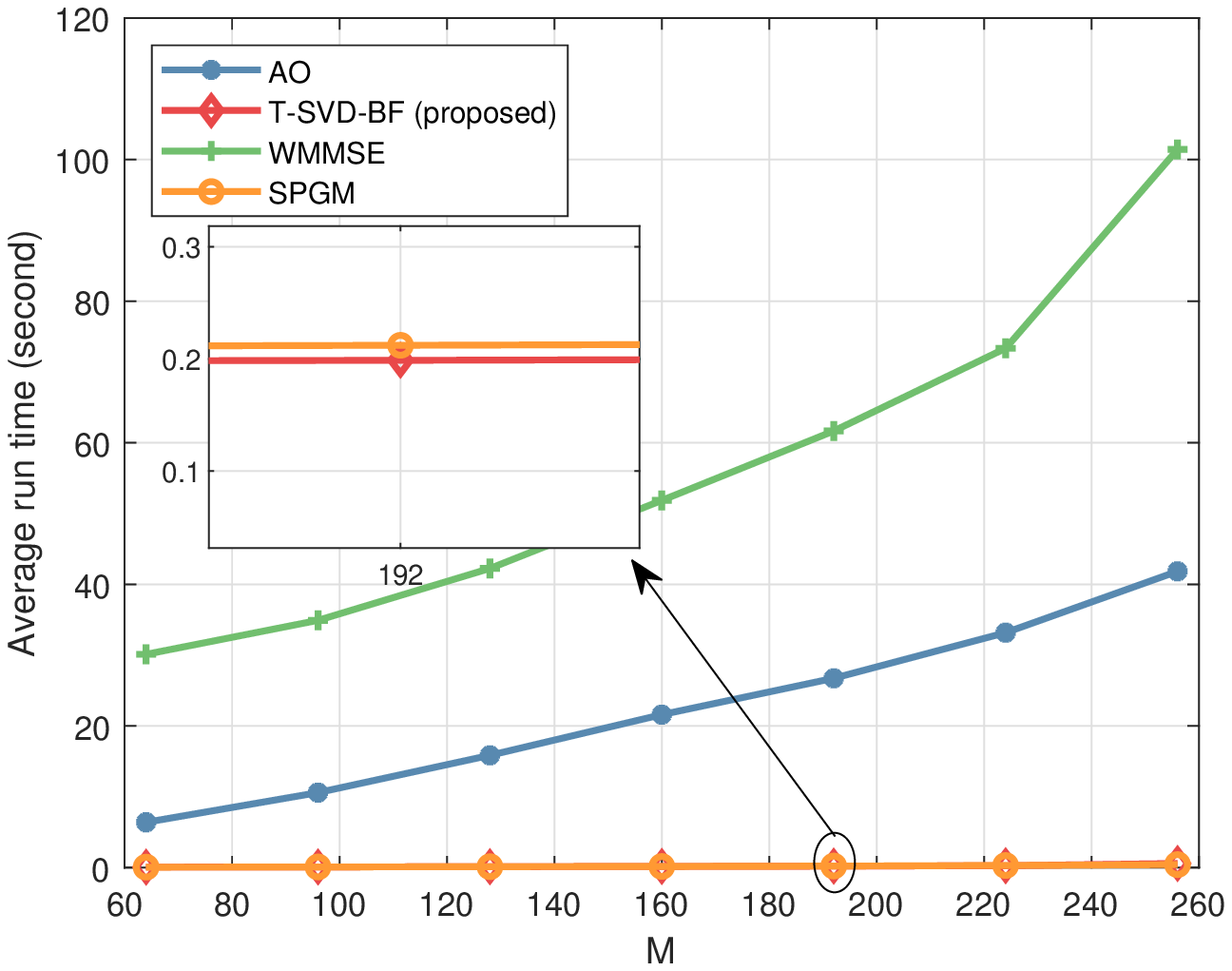}}
\caption{Spectral efficiency and average run time versus number of LIS elements, $M$.}
\label{figSEvM}
\end{figure*}

\subsection{Perfect Channel State Information}
In this section, we assume perfect channel state information (CSI)
is available at the BS. In Fig. \ref{figSEvSNR}, we plot the
spectral efficiency of respective algorithms versus the transmit
power $\rho$, where ``digital'' and ``hybrid'' respectively
represent the spectral efficiency achieved by fully digital
precoding/combining matrices and hybrid precoding/combining
matrices. Here hybrid precoding/combining matrices are obtained
via the manifold optimization-based method developed in Section
\ref{sec-hyb-man}. We observe that for all four algorithms, the
performance of hybrid precoding/combining is very close to that of
fully digital precoding/combining, which corroborates the claim
that hybrid beamforming/combining with a small number of RF chains
can asymptotically approach the performance of fully digital
beamforming/combining for a sufficiently large number of
transceiver antennas. In addition, we see that our proposed method
T-SVD-BF presents a clear performance advantage over the SPGM
method and the WMMSE-based method. Particularly, the substantial
improvement over the SPGM can be attributed to the fact that our
proposed method yields a well-conditioned effective channel matrix
with a smaller condition number, thus creating a more amiable
propagation environment for data transmission. We also note that
our proposed method achieves performance close to the AO-based
method. The convergence behavior of our proposed manifold-based
algorithm used to solve the passive beamforming design problem
\eqref{opt-man2} is depicted in Fig. \ref{fig_converge}, where we
set $\rho = 30$dBm. It can be seen that the proposed
manifold-based algorithm has a relatively fast convergence rate
and attains the maximum objective function value within only ten
iterations.

%A smaller $M$ leads to a less accurate approximation.

%when $M$ is small, say, $M=64$. This performance loss is due to
%the fact that the accuracy of the truncated SVD approximation of
%our proposed method is affected by the value of $M$, as can be
%seen from analysis in Section \ref{sec-dis}.

%when $M$ increases to a moderately large value (say $M=256$), our
%proposed method suffers only about $4\%$ performance loss (about
%$2.4$ bits/s/Hz).

Fig. \ref{figSEvM} depicts the spectral efficiency and average run
time of respective algorithms versus the number of LIS elements,
where the transmit power $\rho$ is set to $50$dBm, and to vary
$M$, we fix $M_y = 16$ and increase $M_z$. All algorithms are
executed on a 2.90GHz Intel Core i7 PC with 64GB RAM. We see that
for all four algorithms, the spectral efficiency increases as the
number of passive elements increases. For this setup, the SPGM and
the WMMSE methods achieve similar performance. Our proposed method
outperforms the SPGM and the WMMSE by a big margin, and the
performance gap becomes more pronounced as $M$ increases. It is
also observed that the spectral efficiency of our proposed method
is slightly lower (by about $4\%$ spectral efficiency loss) than
that of the AO-based method. Nevertheless, our proposed method is
much more computationally efficient than the AO-based method, as
reported in Fig. \ref{figSEvM} (b). For instance, the average run
time required by our proposed method T-SVD-BF is $0.1988$ second
when $M=192$, while it takes the AO-based method about $26.7357$
seconds to yield a satisfactory solution. Such a substantial
complexity reduction makes our proposed method a more practical
choice for LIS-assisted mmWave communications where a large number
of antennas as well as a large number of reflecting elements are
likely to be employed to compensate the severe path loss. In
addition, it is observed that our proposed method outperforms the
WMMSE-based method in terms of both spectral efficiency and
computational complexity.

\begin{figure}[!t]
    \centering
    {\includegraphics[width=3.5in]{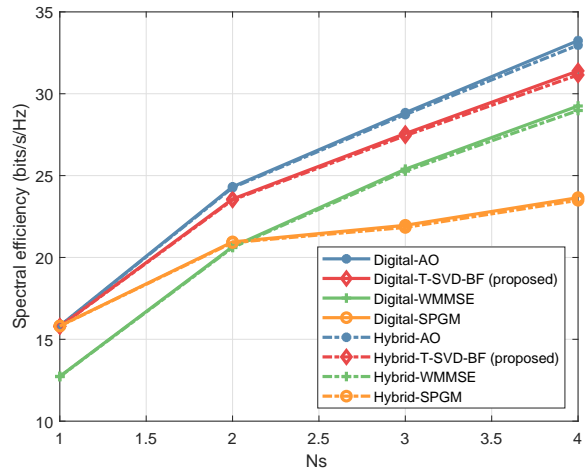}}
    \caption{Spectral efficiency versus the number of data streams $N_s$.} \label{figSEvNs}
\end{figure}

In Fig. \ref{figSEvNs}, we plot the spectral efficiency versus the
number of data streams, where the transmit power $\rho$ is set to
$30$dBm and the number of passive elements is set to $M=256$. It
is observed that the spectral efficiency of T-SVD-BF, AO and WMMSE
improves rapidly as $N_s$ increases, while the performance
improvement of SPGM is quite limited. Again, we observe that
T-SVD-BF achieves a substantial computational complexity
reduction, at the expense of mild performance degradation compared
to the AO-based method. The average run time of our proposed
method is $0.5637$ second when $N_s=3$, while it takes the
AO-based method about $35$ seconds.

%better than WMMSE-based method and

\begin{figure}[!t]
    \centering
    {\includegraphics[width=3.5in]{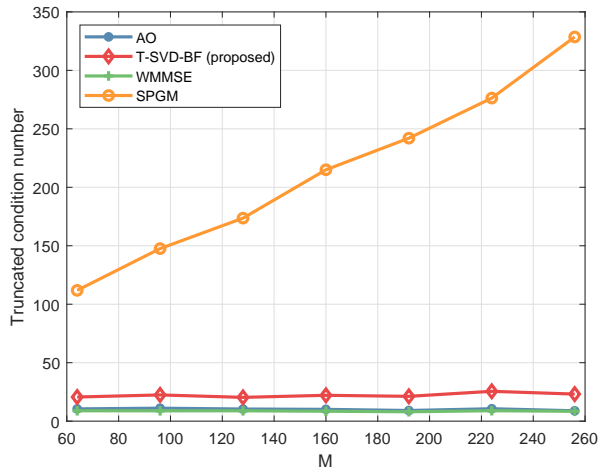}}
    \caption{Truncated condition number versus the number of LIS elements $M$.} \label{figCondivM}
\end{figure}

To gain an insight into how the LIS helps realize a favorable
propagation environment, we use the truncated condition number as
a metric to evaluate the capability of respective algorithms in
reconfiguring the wireless channel. As is well known, the channel
matrix condition number is regarded as an auxiliary metric for
measuring how favorable the channel is. The truncated condition
number is defined as
$\frac{\boldsymbol{\Sigma}_1^2(1,1)}{\boldsymbol{\Sigma}^2_1(N_s,N_s)}$,
where $\boldsymbol{\Sigma}_1(i,i)$ denotes the $i$th largest
singular value of the effective channel
$\boldsymbol{H}_{\text{eff}}$. Fig. \ref{figCondivM} shows the
truncated condition number versus the number of passive reflection
elements. We observe that the truncated condition number of SPGM
grows as $M$ increases, whereas the truncated condition numbers of
the proposed method and the AO-based method are small and remain
almost unchanged with an increasing $M$. This result indicates
that our proposed method and the AO-based method are superior to
SPGM in building a more favorable wireless channel. It also
explains why SPGM does not gain much with the increase of the
number of passive elements. An interesting observation is that the
WMMSE-based method performs worse than our proposed method,
despite of the fact that it has a smaller truncated condition
number. The reason lies in that although the WMMSE-based method
obtains more balanced singular values, the singular values are
generally smaller (in terms of magnitude) than those singular
values obtained by our proposed method, which prevents the WMMSE
method from achieving a higher spectral efficiency.

\begin{figure}[!t]
   \centering
   {\includegraphics[width=3.5in]{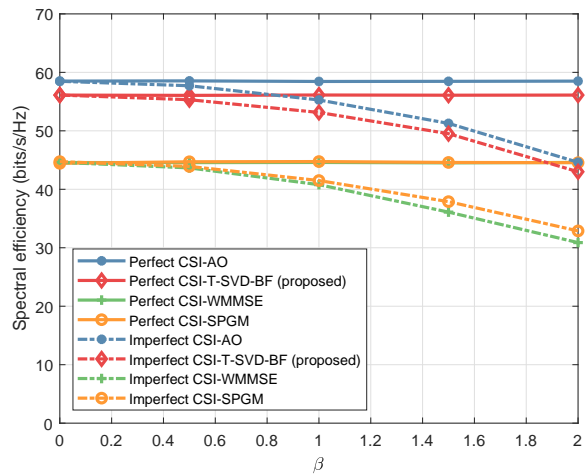}}
   \caption{Spectral efficiency versus the estimation error $\beta$.} \label{figSEvbeta}
\end{figure}

%To investigate the performance loss owing to the beam mismatch (or
%misalignment), which is an important issue in practical, we
%firstly

%where $\beta=\sqrt{3}\Delta$ and $\Delta $ is the standard
%deviation of the beam mismatch error \cite{PradhanLi19}.

\subsection{Imperfect Channel State Information}
In practice, due to channel estimation errors, perfect knowledge
of channel state information (CSI) is usually unavailable. It is
therefore important to evaluate the performance of respective
algorithms with inaccurate knowledge of AoAs and AoDs. To this
end, let $\delta\triangleq\theta-\hat{\theta}$ denote the
estimation error of the AoA/AOD parameter, where $\theta$ denotes
the true AoA (AoD) and $\hat{\theta}$ is the estimated AoA (AoD).
Note that our channel model consists of a set of AoA/AoD
parameters, and we assume that the estimation errors for these
AoA/AoD parameter are modeled as independent and identically
distributed (i.i.d.) random variables following a same uniform
distribution \cite{PradhanLi19}:
\begin{align}
f(\delta)=\left\{\begin{array}{ll}
\frac{1}{2 \beta}, & \text { if }-\beta \leq \delta \leq \beta \\
0, & \text { otherwise }
\end{array}\right.
\end{align}
In our simulations, we set $N_t=N_r=64$, $M = M_y \times M_z = 16
\times 16 = 256$, $L = P = 7$, and $\rho = 50$dBm. Fig.
\ref{figSEvbeta} depicts the spectral efficiency versus the
estimation error $\beta$, where $\beta$ varies from 0 to
$2^{\circ}$, as suggested in \cite{PradhanLi19}. We see that our
proposed method behaves similarly as the AO-based method: both
methods suffer nearly a same amount of performance loss as the
estimation error increases. This result indicates that our
proposed method does not exhibit higher sensitivity to inaccurate
CSI than other methods. Moreover, it can be observed that our
proposed method still presents a clear performance advantage over
the SPGM and the WMMSE-based methods in the presence of channel
estimation errors.

%We see that the the performance degradation owing to the beam
%mismatch is high for all four algorithms when $\beta = 2^{\circ}$.
%When $\beta$ increasing, the performance loss is approaching to
%$0$. Specifically, the proposed T-SVD-BF method suffers nearly
%$26\%$ performance loss while the AO-based method sustains about
%$25\%$ performance degradation when $\beta = 1^{\circ}$.

\section{Conclusion} \label{sec-conclu}
In this paper, we considered an LIS-assisted downlink mmWave MIMO
system with hybrid precoding/combining. The objective is to
maximize the spectral efficiency by jointly optimizing the passive
beamforming matrix at the LIS and the hybrid precoder (combiner)
at the BS (UE). To tackle this non-convex problem, we developed a
manifold optimization (MO)-based algorithm by exploiting the
inherent structure of the effective (cascade) mmWave channel.
Simulation results showed that by carefully designing phase shift
parameters at the LIS, our proposed method can help create a
favorable propagation environment with a small channel matrix
condition number. Besides, it can achieve a performance comparable
to those of state-of-the-art algorithms, while at a much lower
computational complexity.

%Moreover, the AO-based method has a smaller truncated condition
%number than the proposed-MO method, which reveals the reason why
%the proposed-AO based method suffers a slight performance loss.

%It should be note that this metric is different from the
%traditional \emph{condition number}, i.e. $\frac{\Lambda_{\rm
%max}}{\Lambda_{\rm min}}$, where $\Lambda_{\rm max}$
%($\Lambda_{\rm min}$) denotes the maximum (minimum) singular value
%of the effective channel $\boldsymbol{H}_{\rm eff}$. The reason is
%that the number of data streams $N_s$ is limited by the amount of
%RF chains such that $N_s \leq {\rm rank} (\boldsymbol{H}_{\rm eff}
%)$.

\bibliography{newbib}
\bibliographystyle{IEEEtran}

\end{document}